# Quantum Data Encoding and Variational Algorithms: A Framework for Hybrid Quantum–Classical Machine Learning

Bhavna Bose[1] and Saurav Verma[2,*]

[1,2] Department of Information Technology, Mukesh Patel School of Technology Management and Engineering, SVKM's NMIMS

**E-mail**: bhavna.bose@nmims.edu[1] and saurav.verma@nmims.edu[2,*]

## Abstract

The development of quantum computers has been the stimulus that enables the realization of Quantum Machine Learning (QML), an area that integrates the calculational framework of quantum mechanics with the adaptive properties of classical machine learning. This article suggests a broad architecture that allows the connection between classical data pipelines and quantum algorithms, hybrid quantum-classical models emerge as a promising route to scalable and near-term quantum benefit. At the core of this paradigm lies the Classical-Quantum (CQ) paradigm, in which the qubit states of high-dimensional classical data are encoded using sophisticated classical encoding strategies which encode the data in terms of amplitude and angle of rotation, along with superposition mapping. These techniques allow compression of information exponentially into Hilbert space representations, which, together with reduced sample complexity, allows greater feature expressivity. We also examine variational quantum circuits, quantum gates expressed as trainable variables that run with classical optimizers to overcome decoherence, noise, and gate-depth constraints of the existing Noisy Intermediate-Scale Quantum (NISQ) devices. Experimental comparisons with a Quantum Naive Bayes classifier prove that even small quantum circuits can approximate probabilistic inference with competitive accuracy compared to classical benchmarks, and have much better robustness to noisy data distributions. Along with this, the paper also talks about the space of computational complexity, where QML lies in the Bounded-Error Quantum Polynomial Time (BQP) space, and how it can be applied to solve NP-hard problems where such solutions are known to be nonexistent in classical solvers. State-of-the-art toolkits such as IBM Qiskit, Google Cirq, and TensorFlow Quantum support practical realization, and enable reproducible hybrid workflows on superconducting, trapped-ion, and photonic quantum backends. This model does not only explain the algorithmic and architectural design of QML, it also offers a roadmap to the implementation of quantum kernels, variational algorithms, and hybrid feedback loops into practice, including optimization, computer vision, and medical diagnostics. The results support the idea that hybrid architectures with strong data encoding and adaptive error protection are key to moving QML out of theory to practice.

## 1. Introduction

Machine learning has transformed the data analytics industry like never before. Availability of a variety of data in large volumes at high velocity, i.e., 'Big Data', has given the world an opportunity to gain significant insights into human behaviour and helped develop very accurate intelligent systems. As the volume of data increases, so does the difficulty in processing it. There is a need for systems that can efficiently process large data sets, using complex algorithms, in real time and with high accuracy. Quantum computing provides a platform to achieve this, with a few caveats! The principles of Quantum mechanics were introduced in the early 1920s by a group of physicists, including Max Born, Werner Heisenberg, and Wolfgang Pauli, at the University of Göttingen [1-3]. However, Shor's algorithm [4] catapulted the interest in Quantum Computing. Fig. 1 shows the progress of Quantum research from 1982 to 2025. The fast pace of development and interest for different countries and industries suggests that Quantum Computing is the future !

Quantum computing research is advancing at a blinding pace. Google and IBM are the two front-runners in this Quantum race. In just a few of years, IBM has successfully developed a quantum computer capable of running up to 5000 two qubit gate operations. [5]. In December 2024, Google launched their breakthrough 105 Qubit Quantum chip called Willow.[6]. This latest development has tackled a very important challenge in Quantum computing by reducing error and decoherence as the number of Qubits is increased. In early 2025, Microsoft launched Majorana 1, a revolutionary Quantum Processing Unit (QPU) using topological qubit arrays. This breakthrough may lead to the design of millions to qubits on a single chip. [7-8].

As previously mentioned, the principles of Quantum mechanics were known since the early 1920s. However, interest in applications of Quantum Mechanics to Computing grew in the 1990s. Algorithms like Shor's algorithm and Grovers algorithm, used Quantum Computing principles to solve the classical problems of prime number factorization and searching an item in an unordered list respectively with major speedups.

Classical systems store information in a single bit i.e. 0 or 1. In contrast, quantum computers store information in qubits which are represented using the Dirac notation, written as $|\psi\rangle$ (called ket psi). Each single Qubit stores information about two states 0 and 1. On measurement, the Qubit collapses to a single state i.e. 0 or 1. For a 2 qubit system, the possible states are 00,01,10,11. This means that in the intermediate processes (before measurement) fewer qubits (N) can be used to process more (2N) data. All this is possible due to the wave particle duality principle of a quantum mechanics.

This does not mean that classical computers will become obsolete. It is quite the opposite! The current research (as we will see the next section) is working on a symbiotic interaction between Quantum computers and Classical computers. Each processing paradigm has its strength and by harnessing the strength of both systems, solutions of traditionally unsolvable problem can be attempted. The quantum computing offers unique solutions to many unsolvable problems, quantum computing faces its unique challenges. [9] discusses the many challenges faced in the development of quantum computing based systems. These include hardware scalability, decoherence, and limited gate operations to name a few.

### 1.1 Related Literature

The work done in [10-12] discusses the different applications of machine learning, quantum information processing and the overlap between these two fields. For those seeking an entry point into the topics of Quantum Machine Learning (QML), [13] serves as a valuable resource, while [14] emphasizes algorithmic complexity and theoretical foundations through illustrative examples of QML techniques., summarize their characteristics, and offer broader context within this domain. Additionally, [17] demonstrates examples of quantum enhancements to Machine Learning and Artificial Intelligence. Among introductory texts, the monograph by [18] stands out as the most comprehensive resource, presenting significant discoveries in Quantum Machine Learning alongside a thorough introduction to both classical machine learning and quantum computing. Similarly, [19] delves into well-established techniques, offering a detailed overview of the field.

A unique perspective is provided by [20], which explores QML's applications in robotics while also addressing classical QML concepts. Recent works have begun to propose specific application scenarios for QML. For instance, [21] discusses how novel QNN architectures could address numerical challenges in finance, while [22] examines the potential of quantum neural networks to enhance medical image recognition. Physics remains the most prominent area for QML applications, with [23] offering a comprehensive review. Conversely, other cross-disciplinary applications have seen less significant contributions from QML, as noted by [24]. This work aims to give an introduction to Quantum Machine Learning (QML). It covers foundational concepts and proposes a framework for implementing QML. Rather than striving for exhaustive coverage, the goal is to equip readers with sufficient background and guidance for deeper study.

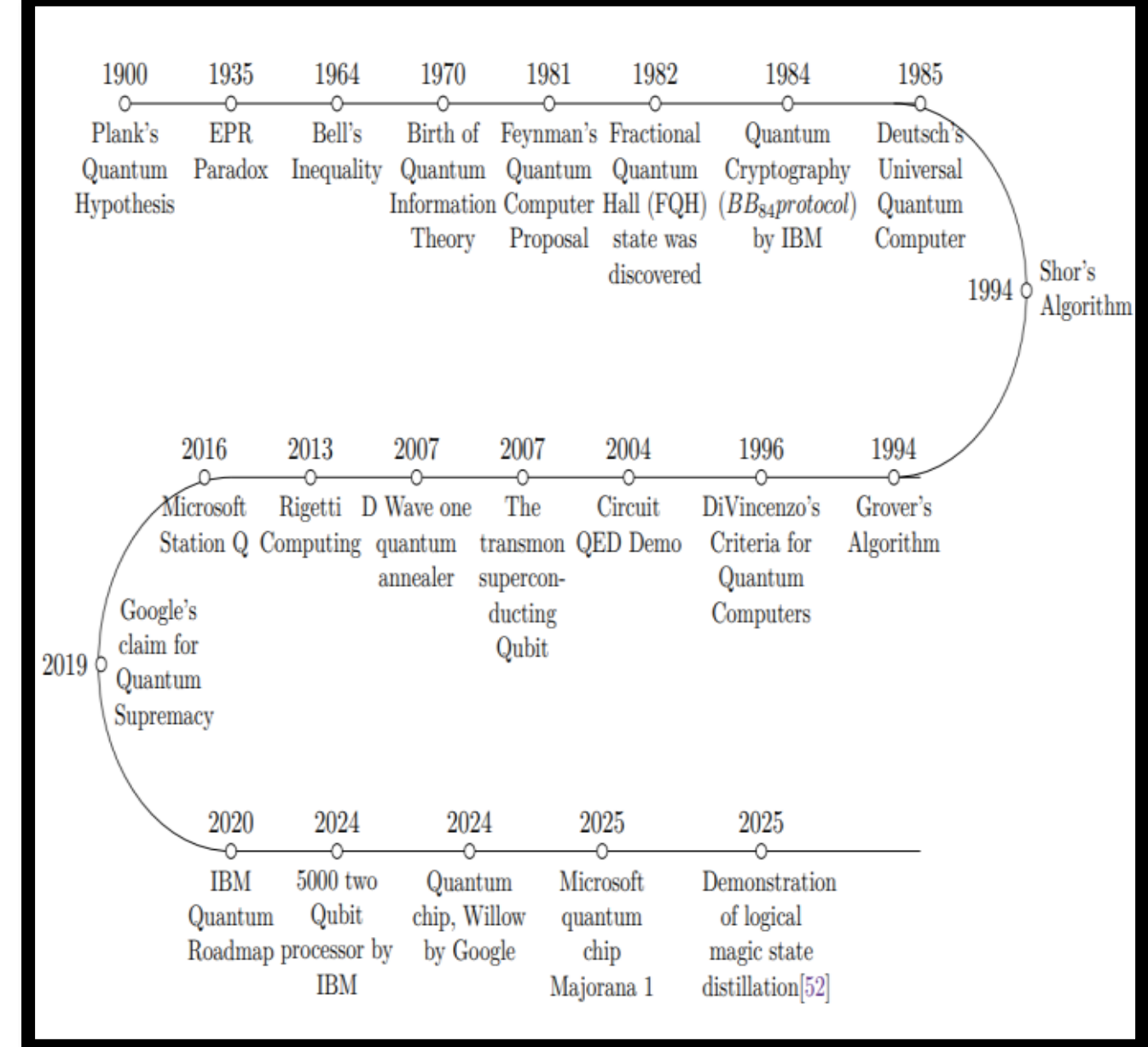

Figure 1. Quantum Computing Timeline: The figure shows the major milestones in the history of quantum computing beginning from the introduction of Quantum theory by Plank to the latest developments in this field.

## 2. Quantum Computing

In this section we will discuss the building block of Quantum systems, Qubits and the types of problems which can be solved using the Quantum approach.

### 2.1 Qubits

Qubits, are the basic building block of a quantum computing system. Qubits in quantum computers are analogous to bits in classical systems. Qubits are quantum particles and can be implemented using many physical particles like superconducting qubits, trapped ion qubits, quantum dots, photons, neutral atoms to name a few. A qubit is represented using the Dirac notation.

**Dirac Notation**

Dirac notation is way of representing Qubits. The Dirac notation consists of the *bra-ket* i.e. $\langle bra|ket\rangle$**.** The represents a column matrix. Bra is the complex conjugate of the ket. Thus bra vector is a row vector. The ket vector contains the probability amplitudes of the qubit for every possible state.

For example:

Consider the Qubit $$|\psi\rangle = \begin{bmatrix}\alpha\\ \beta\end{bmatrix} \qquad (1)$$

*where*

$\alpha^2$ is the probability of ψ measuring to 0 and

$\beta^2$ is the probability of measuring 1. The basis states $|0\rangle$ and $|1\rangle$ are shown below.

$$|0\rangle = \begin{bmatrix}1\\ 0\end{bmatrix} \qquad (2)$$

$$|1\rangle = \begin{bmatrix}0\\ 1\end{bmatrix} \qquad (3)$$

From Eq (1),(2) and (3) we can deduce that,

$$|\psi\rangle = \alpha\,|0\rangle + \beta\,|1\rangle \qquad (4)$$

Here, α and β are complex numbers representing the probability amplitudes of 0 and 1. The state of a Quantum system can be a combination of all the possible states. For example: in a single Qubit system the possible states are $|0\rangle$ and $|1\rangle$. For a two qubit system, the possible states can be $|00\rangle$, $|01\rangle$, $|10\rangle$, $|11\rangle$.

A 2 Qubit quantum state can be represented as,

$$|\varphi\rangle = \begin{bmatrix}\alpha\\ \beta\\ \gamma\\ \delta\end{bmatrix} \qquad (5)$$

$$|\varphi\rangle = \alpha|00\rangle + \beta\,|01\rangle + \gamma\,|10\rangle + \delta\,|11\rangle \qquad (6)$$

Thus,

$$|\varphi\rangle = \alpha\begin{bmatrix}1\\0\\0\\0\end{bmatrix} + \beta\begin{bmatrix}0\\1\\0\\0\end{bmatrix} + \gamma\begin{bmatrix}0\\0\\1\\0\end{bmatrix} + \delta\begin{bmatrix}0\\0\\0\\1\end{bmatrix} \qquad (7)$$

On measurement the result is in any one of $2^n$ state for a n qubit system. The result depends on the square of the probability amplitudes. Referring to Equation 4, the probability of Qubit ψ measuring to 0 is $\alpha^2$ and the probability of measuring 1 is $\beta^2$ . A single qubit can give a value either 0 or 1. Thus, the sum of probability of getting 0 or 1 must be 1. This is shown in the Equation 8.

$$|\,\alpha\,|^2 + |\,\beta\,|^2 = 1 \qquad (8)$$

• **Superposition:**

A Qubit can exist in a combination of state 0 and state 1. This phenomenon is called Superposition. Superposition enables a system to be in a combination of all its possible states. This phenomena can be shown mathematically in Equation 4 and 6.

• **Entanglement**: From Equation 6 we can say that the probability of measuring the first qubit to 0 is $\alpha^2 + \beta^2$. Consequently, the second qubit can be 1 with the probability $\beta^2$and 0 with the probability $\alpha^2$. Entaglement is a state where the value of the second qubit can be gueesed with a 100% probability after measuring the first qubit. Entaglement can be shown mathematically by Equation 9 and 10.

$$|\psi\rangle = \frac{1}{\sqrt{2}}\,|00\rangle + \frac{1}{\sqrt{2}}\,|11\rangle \qquad (9)$$

$$|\psi\rangle = \frac{1}{\sqrt{2}}\,|01\rangle + \frac{1}{\sqrt{2}}\,|01\rangle \qquad (10)$$

We say that the two Qubits are entangled which means even if the two Qubits are located at different geographical locations, their values will always align i.e. if one Qubits measures to 0 we can be sure that the other Qubit will also measure to 0 if they are in the sate represented by Equation 9. If the Qubots are in the state represented by Equation 10, then if one Qubits measures to 0 we can be sure that the other Qubit will also measure to 1.

The different types of qubits used in quantum computing today are listed below:

• **Superconducting qubits**: Superconducting materials like aluminium, niobium and tantalum are used to make the qubits. As extremely low temperatures, they enable fast and fine-tuned computations.

• **Trapped ion qubits**: Qubits implemented as trapped ion particles can stay for longer periods without being disrupted by environmental noise and less errors in measurements.

• **Quantum dots:** they are nano scale semiconductor particles. The electron in a Quantum Dot and confined in a very small three-dimensional space making their energy levels discrete and measurable.
• **Photons:** Photons are light particles which exhibit the phenomena of entanglement and superposition. Also, due to their weak interactions with surrounding environments they more immune to noise.
• **Neutral atoms:** Neutral atoms are atoms with equal number of electrons and protons, making them stable for quantum communications. Quantum information is encoded in the internal states of these atoms. Table 1 gives a summary of the different types of qubits, the companies building Quantum Processing Units (QPUs) using that technology and a few details about each.

Table 1 Qubit technology and companies working on building QPUs using that technology

| **Qubit Technology** | **Industry Leaders** | **Error Rates** | **Pros & Cons** |
|---|---|---|---|
| Superconducting qubits [25-26] | IBM [27], Google [28], Rigetti [29], QuantWare [30] | $\sim 10^{-3}$ for 1-qubit gates; $\sim 10^{-2}$ for 2-qubit gates | **Pros:** Fast gates; Scalable fabrication **Cons:** Requires dilution refrigerators; Significant crosstalk at scale |
| Trapped ion qubits [31-32] | Quantinuum [33], IonQ [34], Alpine Quantum Tech. [35], Oxford Ionics [36] | $\sim 10^{-4}$ (high fidelity) | **Pros:** Long coherence times; High-fidelity gates **Cons:** Slower gate speeds (~μs); Laser stability and precision required |
| Quantum dots [37] | Diraq [38], Quobly [39], Quantum Motion [40] | $\sim 10^{-3}$ to $\sim 10^{-2}$ | **Pros:** Semiconductor-compatible; Small footprint **Cons:** Noise from solid-state environment; Fabrication uniformity |
| Photons [37] | Xanadu [41], ORCA Computing [42], Quantum Computing Inc [43], PsiQuantum [44] | Low decoherence; measurement errors | **Pros:** Room-temperature operation; Long-distance transmission **Cons:** Difficulty in deterministic 2-qubit gates; Photon loss and detection inefficiencies |
| Neutral atoms [45,46] | Pasqal [47], Atom Computing [48], ColdQuanta [49], QuEra [50] | $\sim 10^{-3}$ | **Pros:** Natural scalability (3D arrays); Long coherence **Cons:** Requires precise optical trapping; Laser scalability & error correction |

The spin and amplitude of the qubits carry information. In order to measure and control the spin and amplitude, the qubits have to be kept at very low temperatures, just above absolute zero (-459 degrees Fahrenheit). The main challenge in quantum computing is the difficulty in maintaining such low temperatures. Even with a slight increase in temperature, the quantum particle (Qubit) with change its state i.e. loose its value also called decoherence.[51] Classical bits do not lose their information i.e. they maintain their state over time. However, qubits lose their state very quicky i.e. loose coherence in less than 300μs![52]. Low coherence implies that all computations must be completed before the state of the Qubit cannot be measured.

### 2.2 Classical Vs Quantum Complexity

Problems which can be solved efficiently with the classical computers, do not require to migrate to the more computationally expensive quantum systems. The question which arises is, which kind of problems are apt for quantum computations? Computational complexity of the problems needs to be explored to understand the application area of Quantum Computing. This section will discuss the different computation complexity classes. Computational complexity refers to the time and space required to compute a problem which is normally a function of the number of inputs i.e. input size. Fig. 2 shows the set of decision problems that require a certain amount of time and space.

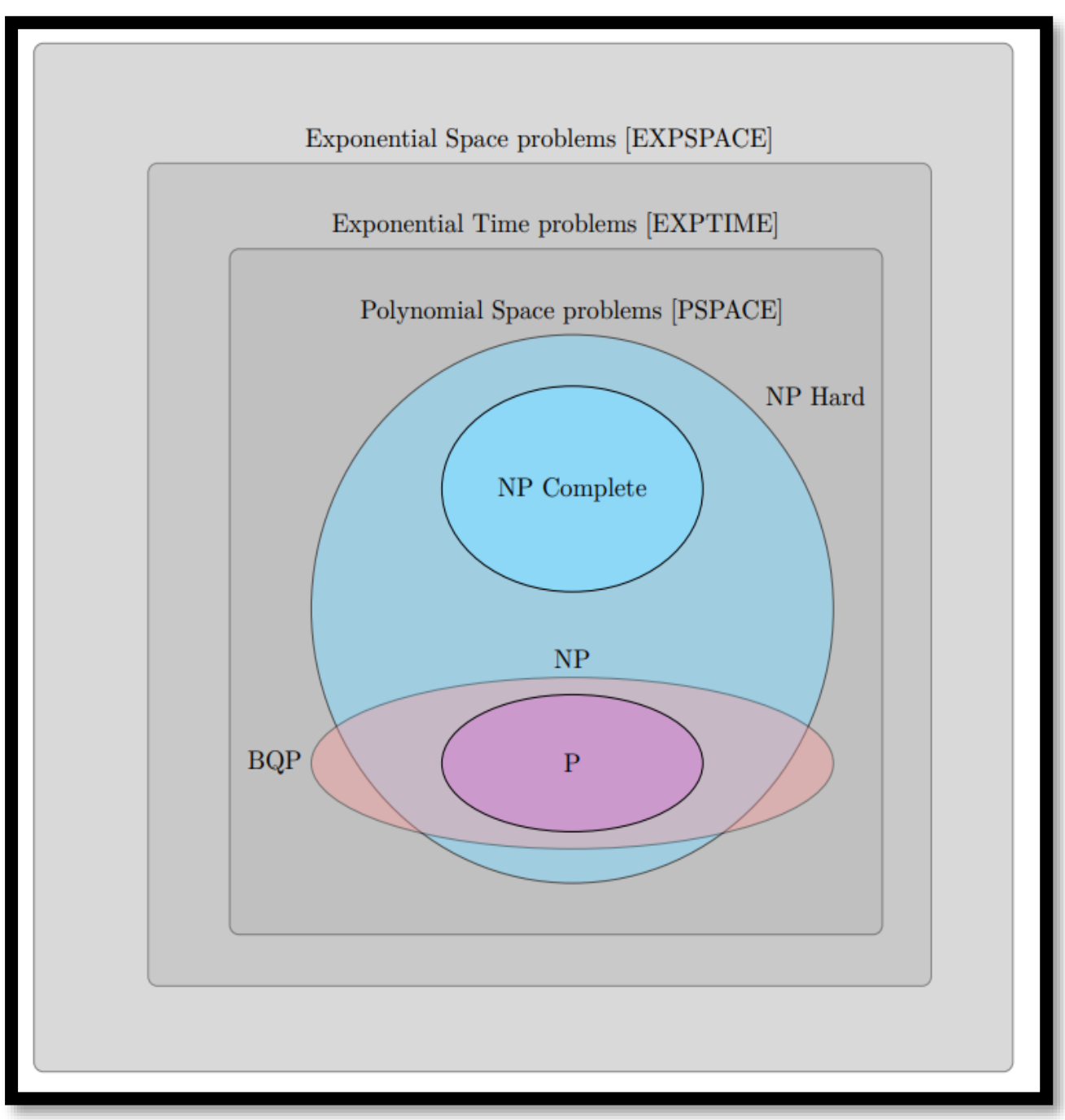

Figure 1 Complexity Classes

- **Exponential Space Problems (EXPSPACE):** The space required to solve problems in this class is an exponential function of the number of inputs.
- **Exponential Time Problems (EXPTIME):** The time required to solve problems in this class is an exponential function of the number of inputs
- **Polynomial Space Problems (PSPACE):** The space required to solve problems in this class is a polynomial function of the number of inputs.
- **Polynomial time problems (P):** Problems in this class can be solved in polynomial time as the input size increases.
- Non deterministic Polynomial time Problems (NP): Problems in the category cannot be solved in polynomial time. However, a solution to these problems can be verified in polynomial time.
- **NP-complete problems:** Problems in this class are the most difficult problems in NP and have no known polynomial solution. This is where famous problems like the traveling salesman and the game Sudoku live.
- **Bounded-error Polynomial Problems (BPP):** which can be solved within some error threshold by a probabilistic classical computer in polynomial time.
- **Bounded-error Quantum Polynomial Problems (BQP):** This is the quantum equivalent of BPP. It is the class of decision problems solvable by a quantum computer in polynomial time with a small chance of error.

## 3. Quantum Logic

Quantum computing paradigm has certain core differences which makes it attractive. Due to the concepts of entanglement, superposition and superdense coding, quantum algorithms provide significant speedups to solve problems considered unsolvable earlier. In an n-Qubit system, there are $2^n$ probability amplitudes, which might suggest an enormous amount of information. The reality is that the measurements we can perform limit how much information we can actually extract. This idea is captured by Holevo's bound, which tells us that n qubits can encode, at most, n bits of information. [53].

### 3.1 Quantum Circuits

Classical computations are performed by applying logic gates to the input bits. The logic gates transform the state of the bits based on certain rules. Analogously, quantum gates are applied to qubits to perform certain computations and give results. Much like classical logic gates, a quantum logic gate transforms, the input A to A′ based on a transformation function applied to it through the logic gate. We usually denote a quantum state as $|\psi\rangle = \alpha\ |0\rangle + \beta\ |1\rangle$. When the quantum logic gate is applied to the quantum input state $|\psi\rangle$, it transforms to $|\psi\ '\ \rangle$. [18]. Listed in Table. 2. are few quantum gates with their symbols and transformation matrices.

Table 2 Quantum Gates, Symbols, Matrices and Resultant Equation

| **Gate Name** | **Symbol** | **Matrix Representation** | **Equation** |
|---|---|---|---|
| **Hadamard** | H | $\frac{1}{\sqrt{2}}\cdot\begin{bmatrix}1 & 1\\1 & -1\end{bmatrix}$ | Input: $\|\psi\rangle = \alpha\ \|0\rangle + \beta\ \|1\rangle$<br>Output: $\|\psi'\rangle = \frac{(\alpha+\beta)}{\sqrt{2}}\cdot\|0\rangle + \frac{(\alpha+\beta)}{\sqrt{2}}\cdot\|1\rangle$ |
| **Pauli-X** | X | $\begin{bmatrix}0 & 1\\1 & 0\end{bmatrix}$ | Input: $\|\psi\rangle = \alpha\ \|0\rangle + \beta\ \|1\rangle$<br>Output: $\|\psi'\rangle = \alpha\cdot\|1\rangle + \beta\cdot\|0\rangle$ |
| **Pauli-Y** | Y | $\begin{bmatrix}0 & -i\\i & 0\end{bmatrix}$ | Input: $\|\psi\rangle = \alpha\ \|0\rangle + \beta\ \|1\rangle$<br>Output: $\|\psi'\rangle = i\alpha\cdot\|1\rangle - i\beta\cdot\|0\rangle$ |
| **Pauli-Z** | Z | $\begin{bmatrix}1 & 0\\0 & -1\end{bmatrix}$ | Input: $\|\psi\rangle = \alpha\ \|0\rangle + \beta\ \|1\rangle$<br>Output: $\|\psi'\rangle = \alpha\cdot\|0\rangle - \beta\cdot\|1\rangle$ |
| **Controlled -NOT (CNOT)** | | $\begin{bmatrix}1 & 0 & 0 & 0\\0 & 1 & 0 & 0\\0 & 0 & 0 & 1\\0 & 0 & 1 & 0\end{bmatrix}$ | Input:<br>$\|\psi'\rangle = \alpha\ \|00\rangle + \beta\ \|01\rangle + \gamma\ \|10\rangle + \delta\ \|11\rangle$<br>Output:<br>$\|\psi'\rangle = \alpha\ \|00\rangle + \beta\ \|01\rangle + \gamma\ \|11\rangle + \delta\ \|10\rangle$ |

### 3.2 Sample Quantum Circuit

A Quantum Circuit is a connection of quantum gates. A quantum circuit contains quantum channels to transmit Qubits. On measurement the Qubit collapses to a classical state. The circuit along with the Qiskit code is illustrated in Fig.3. Fig 3(a) Shows the first step in circuit design i.e. deciding the number of qubits and number of classical bits. In this case, 2 qubits and 2 bits are required. Then a Hadamard gate is applied to make the probability of every outcome equal i.e. to put the Qubits in equal superposition. The resultant circuit for the code to the right is shown to the left. In Figure 3(b) a Controlled NOT gate i.e. CNOT gate is applied to the first Qubit and second qubit with the first qubit as control qubit and second qubit as target. Finally, in Figure 3(c) the two Qubits are measured and their values are recorded in classical bits. The final circuit representation for the code written in Figure 3(a), 3(b) and 3(c) is displayed in Figure 3(d).

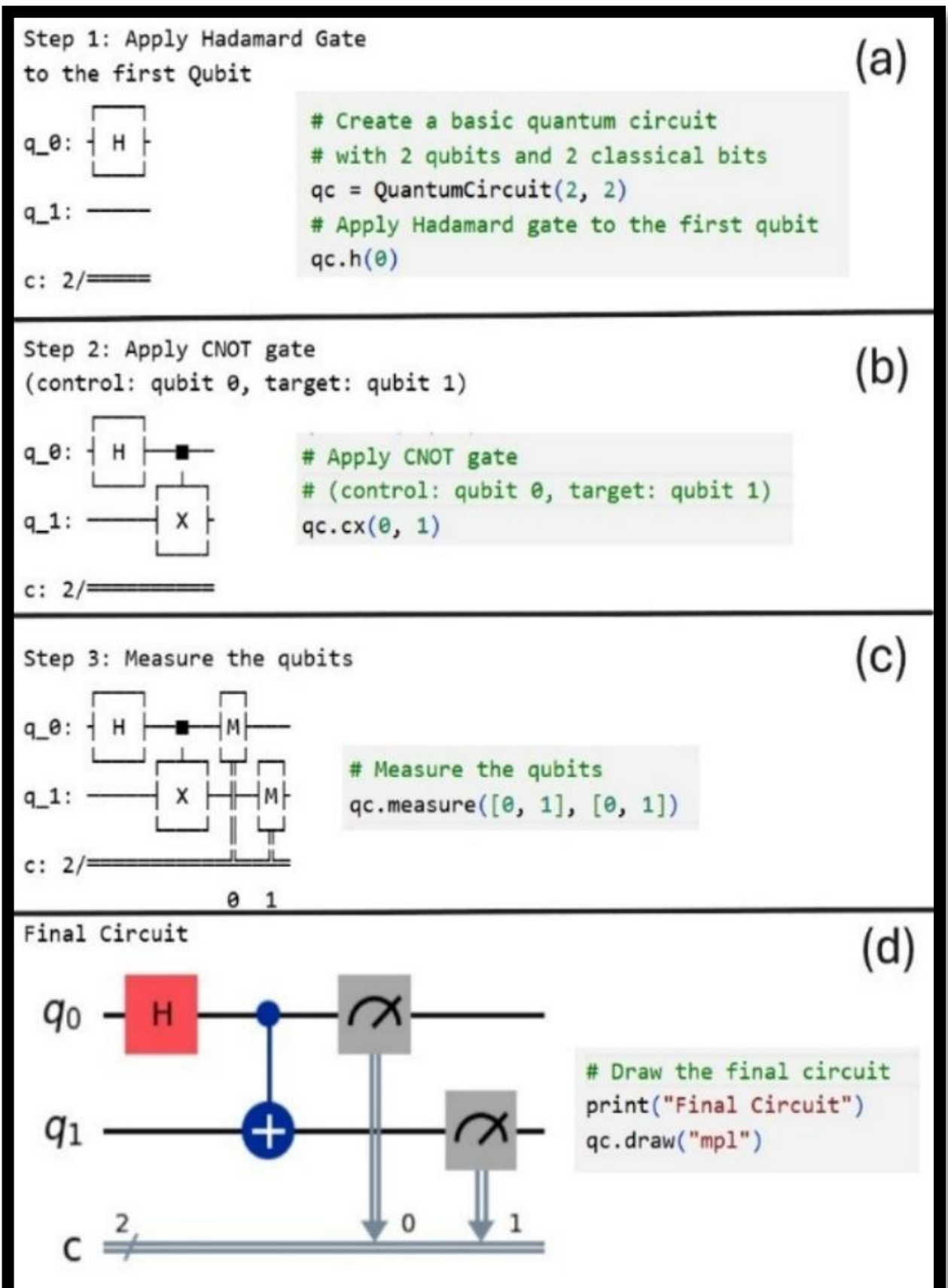

Figure 2 Sample quantum circuit to implement entanglement, along with the code snippet using the Qiskit QuantumCircuit class. (a)Shows the first step in circuit design i.e. deciding the number of qubits and number of classical bits. In this case, 2 qubits and 2 bits are required. Then a Hadamard gate is applied to make the probability of every outcome equal i.e. to put the Qubits in equal superposition. (b) A Controlled NOT gate i.e. CNOT gate is applied to the first Qubit and second qubit with the first qubit as control qubit and second qubit as target. (c) Finally, the two Qubits are measured and their values are recorded in classical bits. (d) The final circuit representation for the code written in (a),(b) and (c).

## 3.2 Quantum Circuit Model

The key components of a quantum circuit as illustrated in Fig. 4 are:

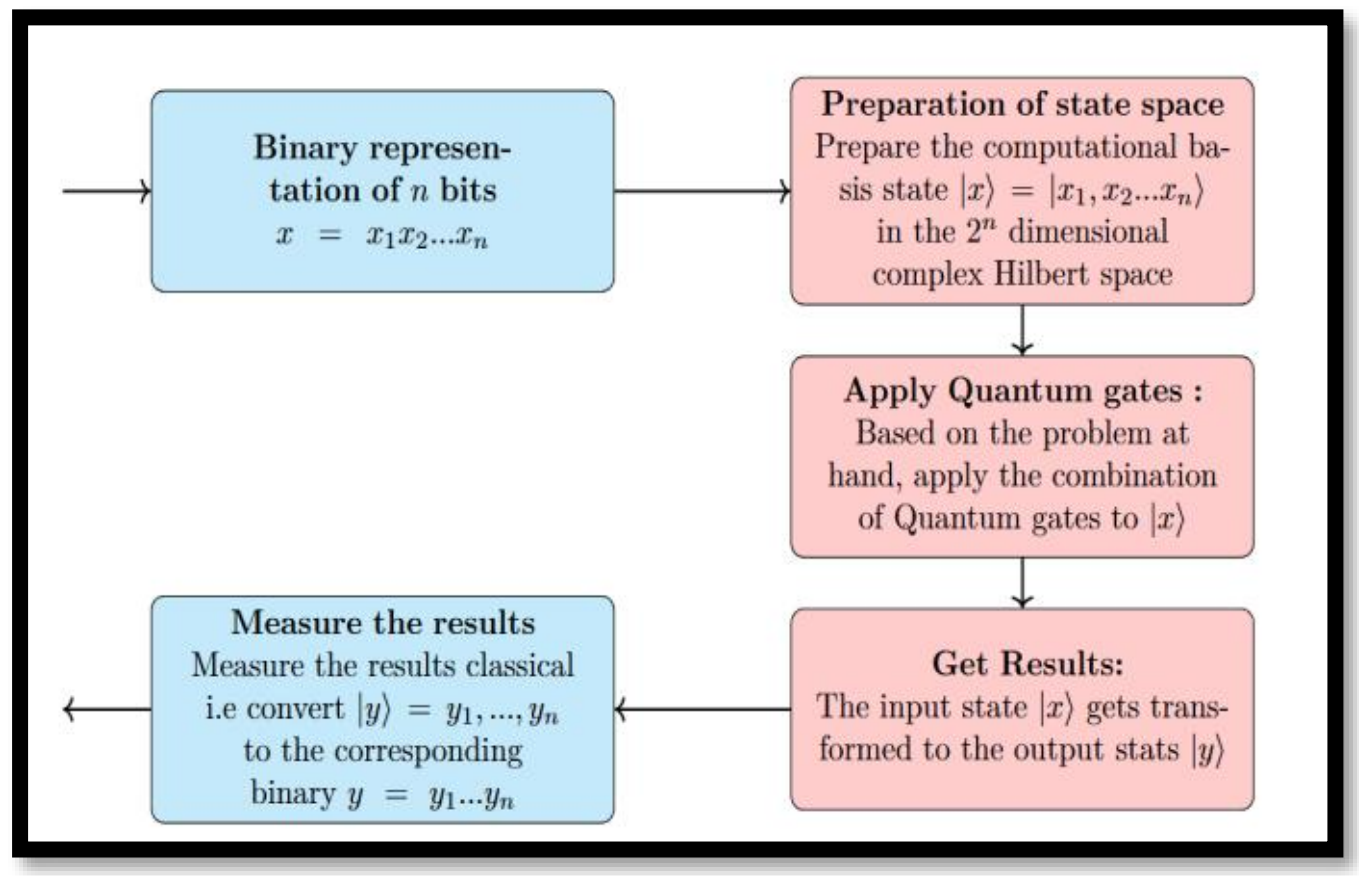

Figure 3 Quantum Circuit Model: The blocks on the left show the classical computation part which includes the binary inputs and measurements of the results in the form of bits. The block on the right show the quantum computation part including blocks to convert the binary inputs to qubits, applying quantum gates for processing and getting results.

**1. Classical Resources**: Quantum circuit can be considered to be divided into two parts: Classical part and Quantum part. The true strength of a quantum system is harnessed by leveraging the strengths of both the computation paradigms and integrating both the systems.

**2. State Space**: Similar to bits in classical systems, quantum computers work on a number of qubits. The qubits occupy a state space, which is a 2n dimensional Hilbert Space, where n is the number of Qubits.

A computational basis state |x⟩ corresponds to the binary number x where

$$|x\rangle = |x_1x_2 \dots x_n\rangle, \qquad (11)$$

$$x = x_1x_2 \dots x_n \text{ and}$$
$$x_i = 0, 1$$

**3. Prepare the state space:** The computational basis states $|x_1x_2 \dots x_n\rangle$ are prepared in at most n steps.

**4. Apply gates and compute:** Once the system is in its initial state, Quantum gates are applied to the required Qubits as per the model/ problem solution and computations are performed. The combination of quantum gates applied is called the Oracle. The Oracle is like the black box which takes the qubits as inputs and transforms them to give results.

**5. Measure the results:** After the computations, the results are measured in the computational basis state $|x_i\rangle$. The results are measured classically as $x_i$ corresponding to $|x_i\rangle$.

## 4. Machine Learning

Machine learning tries to mimic the human nature of learning from past experiences to react to new stimuli. What this means is, based all the past data collected, a model is developed which then gives a prediction of the outcome. The prediction is probabilistic and rarely 100% accurate. Machine learning process generally involves seven steps [54] as shown in Fig. 5. The steps are:

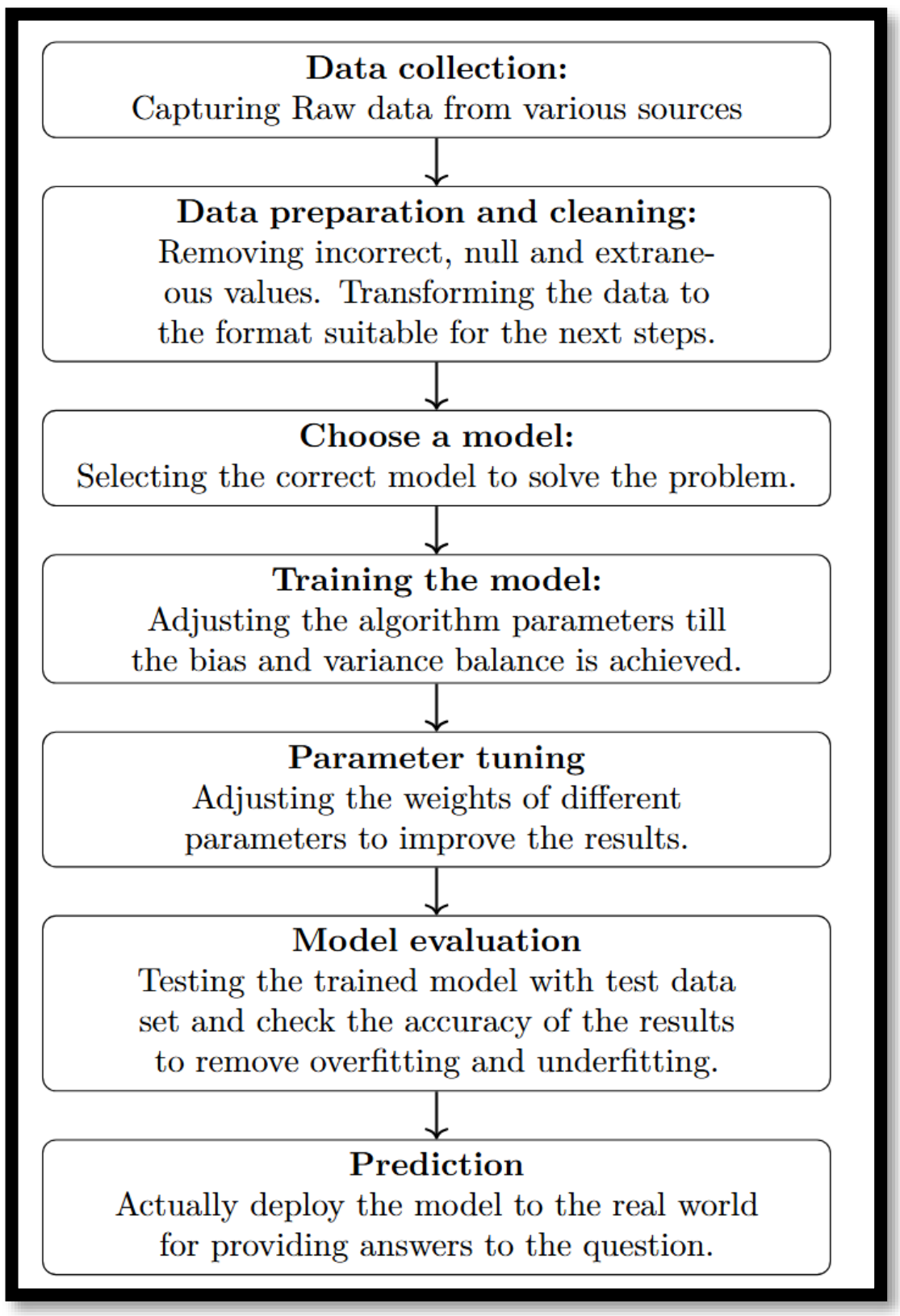

Figure 4 Machine Learning Steps

1. **Data collection:** The raw data is captured from different sources depending on the problem we are trying to solve. For example, if we are trying to detect intruders from CCTV camera footage, the data would be the CCTV camera video footage.
2. **Data preparation and cleaning:** Raw data taken from different sources may be in different formats. There may be missing and invalid/incorrect values. The tasks performed are - Removing incorrect, null and extraneous values and transforming the data to a uniform format suitable for the next steps. This step is important to ensure correct model training and results.
3. **Choose a model:** In this step, the most appropriate machine learning model for the given problem statement is identified.
4. **Training the model:** The data set is divided into training and testing sets. The training data is applied to the model and the algorithm parameters are adjusted till the bias and variance balance is achieved.
5. **Parameter tuning:** The weights are adjusted based in the training data values for optimal results.
6. **Model evaluation:** The values are of the test data set are applied to the trained models and model is evaluated based on parameters like accuracy, F1 score etc to remove overfitting and underfitting.
7. **Prediction:** The model is deployed to the real world for providing answers to the question.

## 5. Quantum Machine Learning (QML)

### 5.1 Introduction to QML

Machine learning relies heavily on probabilities. This is what makes Quantum systems a good candidate for machine learning application as Quantum Computing also, basically relies on calculating the probability of measuring the qubits to a to a particular classical bit. Quantum computers, owing to their strength in handling vast quantities of data and parallel processing [54], can aid in machine learning tasks. The power of Quantum computing can be leveraged in Machine learning to give 4 combinations of approaches: The work done in [55] discusses these four approaches as shown in Table 3:

**Table 2 Different approaches to QML**

| Approach | Concept | Working | Example |
| --- | --- | --- | --- |
| **Classical-Classical (CC)** | Data in the form of bits is captured and analysed by traditional systems that simulate quantum principles. | Runs on classical computers using bits, incorporating quantum-inspired techniques to improve performance. | Algorithms using quantum-inspired optimization (e.g., Quantum Annealing simulators running on classical hardware). |
| **Classical-Quantum (CQ)** | Classical (bit-based) data is encoded into qubits and processed using quantum computers. | Classical data is converted to quantum states, processed by quantum circuits or algorithms, and decoded back into classical results. | A Quantum Neural Network (QNN) where binary inputs are encoded as qubits, processed using quantum gates, and measured to yield |

| | | | classical output. |
|---|---|---|---|
| **Quantum-Classical (QC)** | Classical ML techniques are applied to data originally in qubit form. | Quantum data (e.g., from quantum communication systems) is measured to obtain classical data, which is then analysed using standard ML models. | Applying support vector machines or clustering algorithms to classical measurements obtained from quantum experiments. |
| **Quantum-Quantum (QQ)** | Quantum data is directly processed by quantum algorithms on quantum hardware. | Fully quantum pipeline: both data and computation occur using quantum states, enabling learning from quantum patterns without classical conversion. | Quantum kernel methods or variational quantum algorithms learning directly from entangled quantum states. |

The choice of using the classical approach or quantum approach requires some analysis. As mentioned before, quantum computing algorithms give a clear advantage over classical systems for algorithms with very high sample complexity [19]. Quantum approaches due to their probabilistic nature, are sometimes more robust to noise than classical algorithms. [56, 57]. Some metrics used in model selection are computational complexity, sample complexity, robustness to noise, circuit depth (number of layers), accuracy and optimum bias variance trade-off.[19].
The most common approach for implementing QML is the CQ setting. The two ways to implement QML is the CQ setting are [19]:

a) **The translational approach**: Some parts of the classical algorithms are translated i.e. run using Quantum approaches. The choice of using classical or quantum approach is which approach will give better performance (accuracy or speed wise). Few examples include Variational Quantum Eigensolver, Quantum Machine Learning , Quantum Approximate Optimisation Algorithm etc
b) **The exploratory approach:** The algorithms are run entirely on quantum systems and may not have any quantum counterpart.

Few examples are Shor's Algorithms, Quantum Walks etc

### 5.2 Quantum Data Encoding

The work done in [58-59] discuss the quantum data encoding methods used to prepare the data to be input to a machine learning algorithm. The three approaches discussed are:

#### 5.2.1 Basis Encoding

In this method the qubits are directly mapped to the bit values i.e. bit 0 and 1 will be represented as $|0\rangle$ and $|1\rangle$ respectively. A classical number 110 can be represented as the quantum state $|110\rangle$ as shown in Equation 12.

$$|110\rangle = \alpha\,|1\rangle \otimes \beta\,|1\rangle \otimes \gamma\,|0\rangle\,, \qquad (12)$$

here α, β, and γ are the information stored in the form of probability amplitudes for each Qubit state.
This encoding strategy very simple to use and utilises n qubits to represent a n bits binary number.

#### 5.2.2 Superposition encoding

In this encoding scheme, the Qubit state is encoded as a superposition of the basis states.

$$|110\rangle = \frac{1}{\sqrt{3}}\,|100\rangle + \frac{1}{\sqrt{3}}|010\rangle + \frac{1}{\sqrt{3}}|001\rangle \qquad (13)$$

$$|78\rangle = \frac{1}{\sqrt{2}}\,|1001110\rangle + \frac{1}{\sqrt{2}}|0100111\rangle \qquad (14)$$

Superposition encoding uses the strength of Quantum parallelism enabling one operation to be performed on all the many inputs.

#### 5.2.3 Angle encoding

When the Qubits are represented on the Bloch sphere, the probabilities of basis states are the phase shift from the X, Y and Z axes. Thus, in angle encoding, the classical data is represented as a phase shift. Quantum gates use rotation operations around different axes (X, Y, or Z) for changing the quantum state.
The rotation around the X, Y and Z axis denoted as $R_x(\theta)$, $R_y(\theta)$ and $R_y(\theta)$ respectively and is shown in Equations 8,9 and 10.

$$R_x(\theta) = e^{-i\theta X/2} = \begin{bmatrix} Cos\,(\frac{\theta}{2}) & -i\,Sin\,(\frac{\theta}{2}) \\ -i\,Sin\,(\frac{\theta}{2}) & Cos\,(\frac{\theta}{2}) \end{bmatrix} \qquad (15)$$

$$R_y(\theta) = e^{-i\theta Y/2} = \begin{bmatrix} Cos\,(\frac{\theta}{2}) & -Sin\,(\frac{\theta}{2}) \\ Sin\,(\frac{\theta}{2}) & Cos\,(\frac{\theta}{2}) \end{bmatrix} \qquad (16)$$

$$R_z(\theta) = e^{-i\theta Z/2} = \begin{bmatrix} e^{-i\theta/2} & 0) \\ 0 & e^{i\theta/2} \end{bmatrix} \quad (17)$$

### 5.2.4 Amplitude encoding

In this scheme, classical information is encoded in the probability amplitudes of quantum state.If the number of values to be represented is n, then the number of qubits required are $\log_2(n)$.The values of probabilities of each Qubit state are the square root of the data value divided by the square root of the sum of squares of all the n data values. The quantum state can be represented as:

$$|\psi(X)\rangle = \sum_{i=1}^{n} x_i |i\rangle \quad (18)$$

where $x_i$ is the $i^{th}$ element of the vector X, and $|i\rangle$ denotes the computational basis states of the qubits. For this example, n = 4, and the coefficients xi correspond to the values [1.2, 2.7, 1.1, 0.5].Each term is square-normalized by dividing it by the square root of the sum of squares of all elements. For the vector X = [1.2, 2.7, 1.1, 0.5], the normalization factor is:

$$\sqrt{\sum_{i=1}^{n} x_i^2} = \sqrt{1.2^2 + 2.7^2 + 1.1^2 + 0.5^2} = \sqrt{10.19} \quad (19)$$

The normalized quantum state is:

$$|\psi\rangle = \frac{\sqrt{1.2}}{\sqrt{10.19}}|00\rangle + \frac{\sqrt{2.7}}{\sqrt{10.19}}|01\rangle + \frac{\sqrt{1.1}}{\sqrt{10.19}}|10\rangle + \frac{\sqrt{0.5}}{\sqrt{10.19}}|11\rangle \quad (20)$$

In this encoding strategy, n qubits are needed to represent 2n input values.

## 6. Implementation of QML Algorithms

In their work [60], the authors have suggested a framework to implement Hybrid quantum-classical algorithms. As discussed in Table 3, Quantum computing can work in tandem with a classical system in 4 ways i.e. CC, QC, CQ and QQ.

The most common approach is the CQ approach where classical data is encoded into Qubits and processed using quantum computing techniques. The results are then decoded from qubits to classical bits. This approach contains three parts: the classical part, quantum part and the interface between the classical and quantum systems as shown in Fig.6. The data encoding, application of Oracle, amplitude adjustment and optimization steps are repeated multiple times in the range of thousands of iterations to give reliable results. The quantum part comprises of mapping the algorithm to an actual circuit with quantum gates. The steps involved in this process, are shown on the left side of Fig. 6 in red. In the classical component, classical values obtained are optimized using classical techniques. The steps involved are shown on the right-hand side of Fig. 6 in cyan. The Quantum part and Classical part represent data as qubits and bits. At the interface of the two strategies, the data need to be encoded and decoded from classical to quantum and vice versa. The steps are as follows:

1. **(Quantum Part)** Based on the encoding strategy and the input data, the number of qubits required is decided. All the Qubits are initialised to the basis state of $|0\rangle$.
2. **(Quantum Part)** A Hadamard gate is applied to all the qubits to create a uniform superposition. This ensures equal probability of measuring every possible state initially i.e. before processing the inputs..
3. **(Interface Part Classical to Quantum)** Based on the encoding strategy, the classical data is translated to qubits for further processing.
4. **(Quantum Part)** Apply the different Quantum gates to implement the Oracle function $U_f$. The Oracle function will generate a qubit state which is stored in few or all of the Qubits.
5. **(Quantum Part)** The amplitude adjustment part is very crucial to quantum computations. Since, the input qubits are in a superpositions of all possible states, after evaluation by the Oracle, the resultant qubits are also in a superposition of all the states. The square of the amplitudes(i.e. coefficient of the qubits states) is the probability of measuring to that particular state. The goal of this step is to increase the likelihood of arriving at the correct answer. This is achieved by increasing the amplitudes of the qubits with the desired result and reducing the amplitudes of all the others. Since the sum of squares of amplitudes must be equal to one, increase in one amplitude will require a reduction of the amplitude of the others.
6. **(Interface Part Quantum to Classical)** Upon measurement, the Qubit collapses to one of the $2^n$ classical values where n is number of qubits used. The values of the classical bit can be any of the $2^n$ superimposed states. If the computation is repeated many times and the value with maximum occurrence is considered the final result, then the state with the highest probability (i.e. the correct answer) will have maximum occurrence due to the amplitude adjustment step.
7. **(Classical part)** Based on the results obtained, classical optimisation techniques are applied on the data and iterated again for Quantum evaluations.

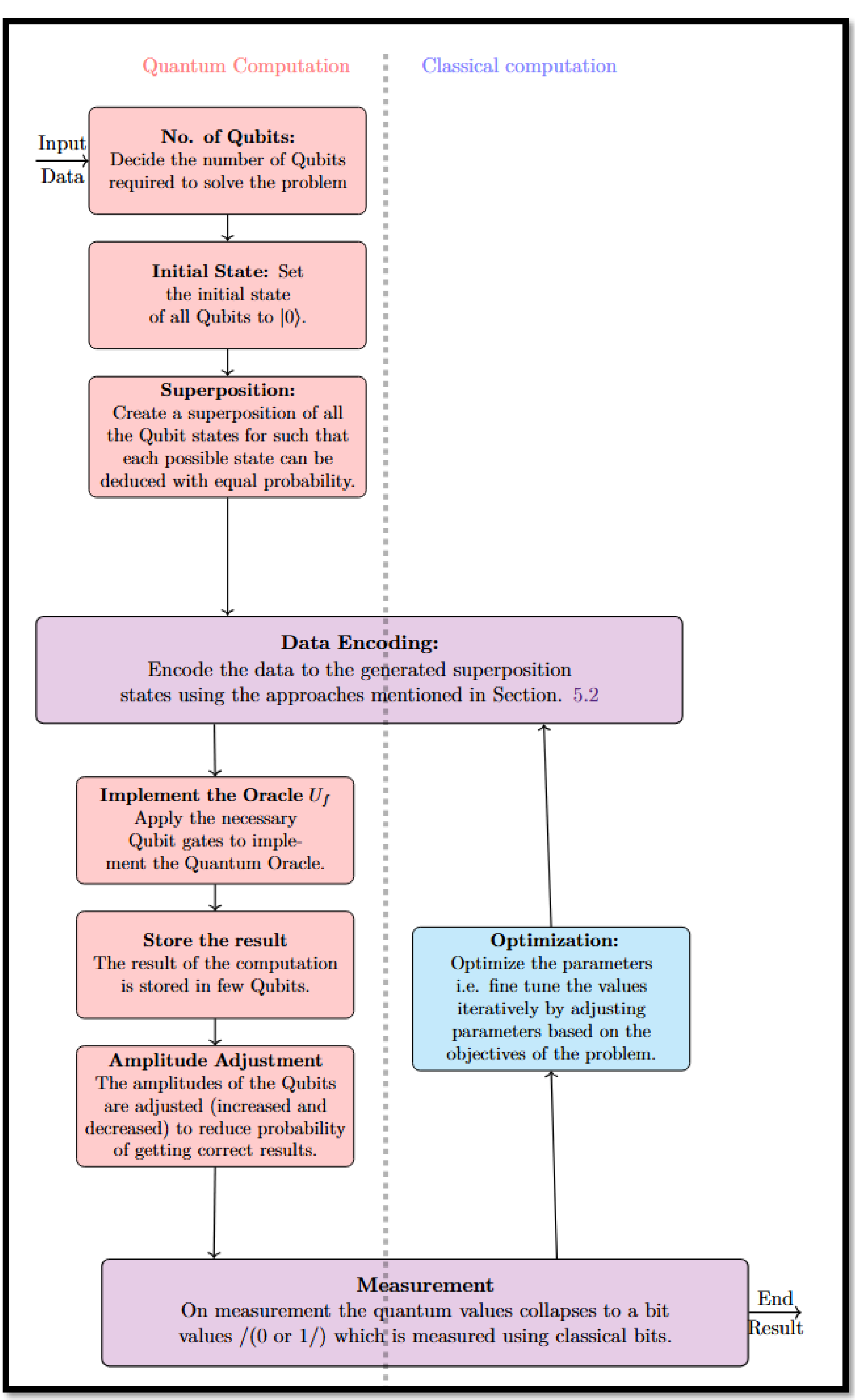

Figure 6 Framework to implement the Quantum Machine learning.

### 6.1 Implementation of QML (Titanic Dataset)

We tested Machine Learning using Naïve Bayes and Quantum Machine Learning with Naïve Bayes on the Titanic Data Set. The steps for both are listed:

**Steps in the Classical Naive Bayes Model (Titanic Dataset)**

**Algorithm 1:** NAIVE BAYES CLASSIFIER FOR SURVIVAL PREDICTION

**Input:** Dataset *D* containing features *{Pclass, Sex}* and label *{Survived}*

**Output:** Trained Naive Bayes model *NB_model,*predictions and evaluation metrics

**Phase 1 –** Training

1. Load dataset *D*
2. Encode categorical variable *Sex* into numerical format {0, 1}.
3. Split D into training set *D_train* and testing set *D_test*.
4. Fit Naive Bayes model *NB_model* using *D_train*.
5. Return *NB_model*.

**Phase 2 -** Prediction

1. **For each** test instance *X* in $D_{test}$ compute:

   *P(Survive|X) =P(Survive)×P(Pclass|Survive)×P(Sex|Survive)* *(21)*

   *P(Dead|X) =P(Dead)×P(Pclass|Dead)×P(Sex|Dead)* ) *(22)*

2. Select the class label (0 or 1) corresponding to the higher posterior probability.
3. Compare predicted labels with ground truth to compute
   - a) Classification accuracy
   - b) Confusion matrix

**Steps in Single-Qubit Quantum Naive Bayes Model**

**Algorithm 2:** SINGLE-QUBIT QUANTUM NAIVE BAYES MODE

**Preprocessing Phase (Classical Part)**

**Input:** Dataset *D* containing features *{Pclass, Sex}* and label *{Survived}*

**Output:** $P_{Survive}$, $P_{Dead}$ for each passenger

1. Load dataset *D* (*Pclass, Sex, Survived*).
2. Compute prior survival probability = fraction of survivors.

$$P(Survive) = \frac{Number\ of\ Survivors}{Total\ Passengers}, M_{pclass} = \frac{P(pclass|Survive)}{P(pclass)}, \quad (23)$$

3. Compute modifiers (likelihood ratios) for *Pclass* & *Sex*:

$$M_{pclass} = \frac{P(pclass|Survive)}{P(pclass)} \quad (24)$$

$$M_{sex} = \frac{P(sex|Survive)}{P(sex)} \quad (24)$$

4. For each passenger, compute combined probability:

$$P_{survive} = P(Survive)\ \boldsymbol{X}\ M_{pclass}\ X\ M_{sex} \quad (25)$$

$$P_{dead} = P(dead)\ \boldsymbol{X}\ (\mathbf{2} -\ M_{pclass})\ X\ (2 -\ M_{sex}) \quad (27)$$

**Prediction Phase (Quantum Part)**

**Input:** $P_{Survive}$, $P_{Dead}$ *for each passenger*

**Output:** *Trained Naive Bayes model* NB_model,*predictions and evaluation metrics.*

1. Convert each probability into a rotation angle:

$$\theta = 2\ Sin^{-1}(\sqrt{P}) \quad (28)$$

2. Build two circuits (one for survival, one for death):
   - **a.** Initialize $|0\rangle$
   - **b.** Apply $R_y(\theta)$ with the calculated probability
   - **c.** Measure qubit (probability of $|1\rangle$ = encoded survival probability)
3. **Run** both circuits using a sampler to obtain probabilities of outcome $|1\rangle$.

   **if** $p_{surv} > p_{dead}$

   Predict "Survive"

   **else**

   Predict "Dead"

   **end**

4. **Compare** predictions with true labels to compute:
   - **a)** Classification accuracy
   - **b)** Confusion matrix

The comparative confusion matrix and accuracy of the Naïve Bayes implementation along with the Quantum Circuit is shown in Figure 7.

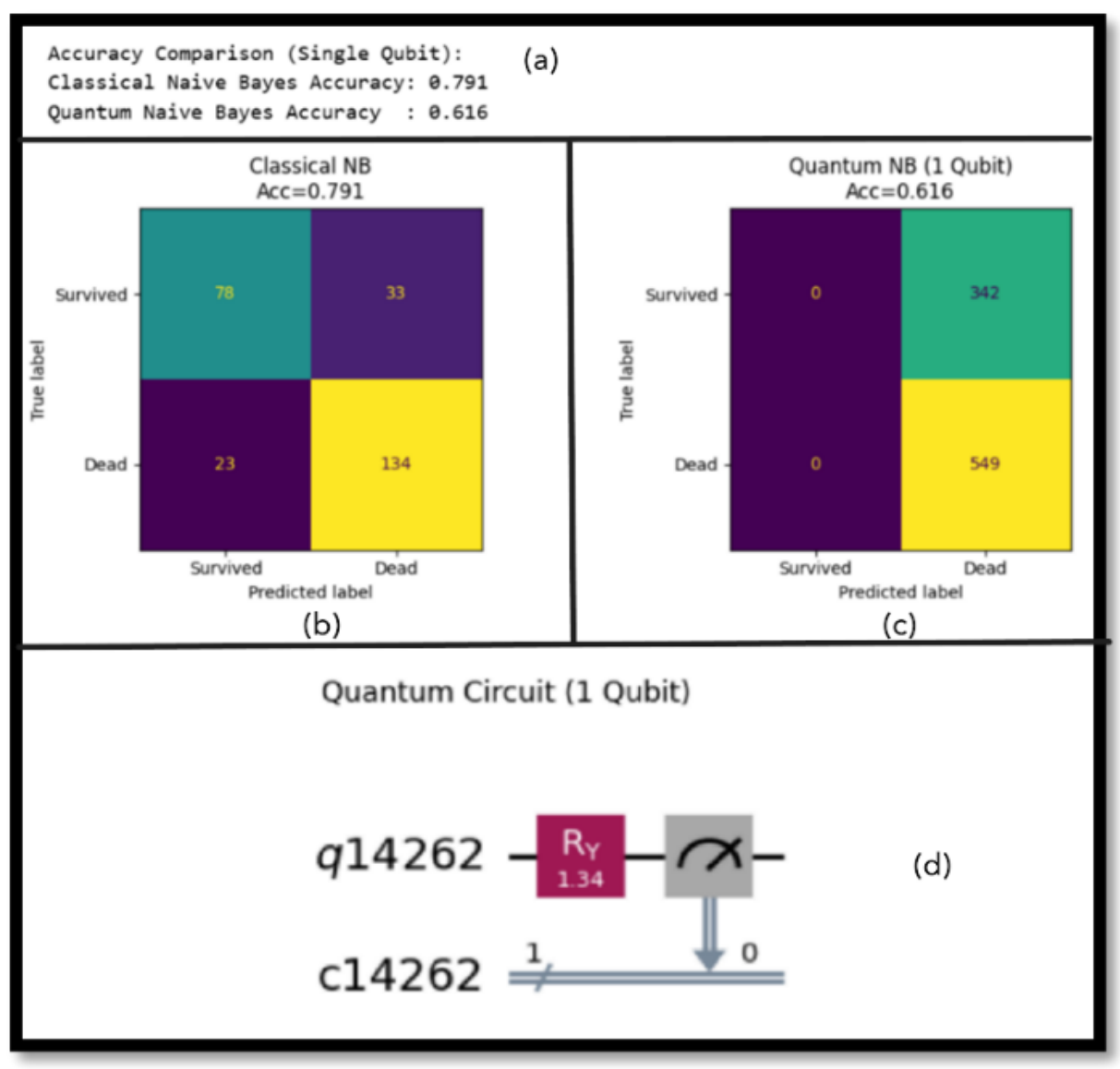

Figure 5 Comparative results of Naive Bayes implementation using Classical ML and QML.

## 7. Tools available for Quantum Computing

Quantum computing has opened doors to solving the problems earlier considered unsolvable. This also poses a threat to our existing system security. All this has sparked a keen interest in Quantum research. A number of tools are available to develop and run quantum algorithms. Some of the tools are tabulated in Table 4.

Table 3 Quantum Development Toolkits

| **SDK / Library** | **Language** | **Hardware** | **Platform / Cloud Service** |
|---|---|---|---|
| IBM Qiskit [61] | Python | IBM Quantum Hardware (Superconducting Qubits) | IBM Quantum Experience |
| Google Cirq [62] | Python | Google's quantum hardware | Google Quantum AI |
| Ms QDK [63] | Q# | Supports multiple hardware providers through Azure | Azure Quantum |
| Rigetti Forest [64] | PyQuil | Rigetti's superconducting quantum processors | Rigetti Quantum Cloud |
| Tensor Flow Quantum [65] | Python | Works on quantum hardware using the Cirq library | TensorFlow Ecosystem |
| Xanadu PennyLane [66] | Python | Xanadu's photonic processors | Xanadu Cloud |
| D-Wave Ocean SDK [67] | Python | D-Wave quantum computers | Leap Quantum Cloud |
| Open Fermion [68] | Python Compatible with Cirq, | Qiskit and other frameworks | Not platform-specific |
| Amazon Braket [69] | Python | Multiple QPUs and simulators | Amazon Web Services |

## 8 Conclusion

Quantum computing is more than just a breakthrough in technology—it's a whole new way of thinking about how we solve problems. By harnessing the strange yet powerful properties of quantum mechanics, quantum computers can tackle challenges that classical systems simply can't handle. From optimizing complex systems to analysing massive datasets, the potential impact on industries like healthcare, finance, and artificial intelligence is staggering. But this isn't a technology we'll see in full force tomorrow. Building practical quantum computers is hard. Qubits, the building blocks of quantum systems, are delicate and prone to errors. Scaling up to larger, more reliable systems will take time and innovation. Even so, companies like IBM and Google have already made impressive strides with powerful quantum processors. Meanwhile, cloud platforms like Amazon Braket and Microsoft Azure Quantum are making quantum computing tools accessible to more people, speeding up the pace of discovery and development. In machine learning, quantum computing offers unique ways to improve existing algorithms and solve new types of problems. Hybrid systems—where quantum and classical computers work together—are showing how these two worlds can complement each other. Libraries and frameworks, like Qiskit and TensorFlow Quantum, are making it easier for developers and researchers to experiment with quantum ideas in practical ways. The potential of quantum computing is undeniable despite the numerous challenges. The future of quantum computing will depend not only on solving technical problems but also on collaborations across science, industry, and policymaking. It's a journey, but one filled with incredible promise. If progress continues at its current pace, quantum computing could soon transform the way we live, work, and think about technology.

## References

[1] Born, M., & Jordan, P. (1925). Zur Quantenmechanik. *Zeitschrift für Physik, 34*, 858–888. https://www.uni-tuebingen.de/ (Retrieved May 29, 2025, via Qwant: Universität Tübingen)

[2] Born, M. (1978). *My life: Recollections of a Nobel laureate*. London: Taylor & Francis.

[3] Broughton, M., *et al.* (2020). *TensorFlow Quantum: A software framework for quantum machine learning* (arXiv:2003.02989). https://arxiv.org/abs/2003.02989

[4] Shor, P. W. (1994). Algorithms for quantum computation: Discrete logarithms and factoring. In *Proceedings of the 35th Annual Symposium on Foundations of Computer Science* (pp. 124–134). IEEE. https://doi.org/10.1109/SFCS.1994.365700

[5] Mandelbaum, R., & Gambetta, J. (2024). IBM Quantum delivers on 2022 100×100 performance challenge | IBM Quantum Computing Blog — ibm.com. *IBM Quantum Blog*. https://www.ibm.com/quantum/blog/qdc-2024

[6] Acharya, R., *et al.* (2024). Quantum error correction below the surface code threshold. *Nature*. https://doi.org/10.1038/s41586-024-08449-y

[7] Aasen, D., *et al.* (2025). *Roadmap to fault-tolerant quantum computation using topological qubit arrays* (arXiv:2502.12252). arXiv. https://arxiv.org/abs/2502.12252

[8] Microsoft Azure Quantum, Aghaee, M., Alcaraz Ramirez, A., *et al.* (2025). Interferometric single-shot parity measurement in InAs–Al hybrid devices. *Nature, 638*, 651–655. https://doi.org/10.1038/s41586-024-08445-2

[9] Sood, S. K., & Pooja. (2024). Quantum computing review: A decade of research. *IEEE Transactions on Engineering Management, 71*, 6662–6676. https://doi.org/10.1109/TEM.2023.3284689

[10] Biamonte, J., *et al.* (2017). Quantum machine learning. *Nature, 549*(7671), 195–202. https://doi.org/10.1038/nature23474

[11] Mishra, N., *et al.* (2021). Quantum machine learning: A review and current status. In *Data Management, Analytics and Innovation: Proceedings of ICDMAI 2020* (Vol. 2, pp. 101–145). Springer.

[12] Schuld, M., *et al.* (2015). An introduction to quantum machine learning. *Contemporary Physics, 56*(2), 172–185. https://doi.org/10.1080/00107514.2014.964942

[13] Dunjko, V., & Wittek, P. (2020). A non-review of quantum machine learning: Trends and explorations. *Quantum Views, 4*, 32. https://doi.org/10.22331/qv-2020-03-17-32

[14] Ciliberto, C., *et al.* (2018). Quantum machine learning: A classical perspective. *Proceedings of the Royal Society A: Mathematical, Physical and Engineering Sciences, 474*(2209), 20170551. https://doi.org/10.1098/rspa.2017.0551

[15] Chakraborty, S., *et al.* (2020). An analytical review of quantum neural network models and relevant research. In *2020 5th International Conference on Communication and Electronics Systems (IC-CES)* (pp. 1395–1400). IEEE. https://doi.org/10.1109/ICCES48766.2020.9137960

[16] Kamruzzaman, A., *et al.* (2020). Quantum deep learning neural networks. In *Advances in Information and Communication: Proceedings of the 2019 Future of Information and Communication Conference (FICC), Volume 2* (pp. 299–311). Springer. https://doi.org/10.1007/978-3-030-12385-7_24

[17] Dunjko, V., & Briegel, H. J. (2018). Machine learning & artificial intelligence in the quantum domain: A review of recent progress. *Reports on Progress in Physics, 81*(7), 074001. https://doi.org/10.1088/1361-6633/aab406

[18] Wittek, P. (2014). *Quantum machine learning: What quantum computing means to data mining*. Academic Press. https://doi.org/10.1016/C2013-0-18854-6

[19] Schuld, M., & Petruccione, F. (2018). *Supervised learning with quantum computers* (1st ed.). Springer Publishing Company. https://doi.org/10.1007/978-3-319-96424-9

[20] Tandon, P., *et al.* (2017). *Quantum robotics: A primer on current science and future perspectives*. Springer. https://doi.org/10.2200/S00746ED1V01Y201612QMC010

[21] Sakuma, T. (2020). Application of deep quantum neural networks to finance. *arXiv preprint arXiv:2011.07319.* https://arxiv.org/abs/2011.07319

[22] Dong, Y., *et al.* (2020). Recognition of pneumonia image based on improved quantum neural network. *IEEE Access, 8*, 224500–224512. https://doi.org/10.1109/ACCESS.2020.3044697

[23] Carleo, G., *et al.* (2019). Machine learning and the physical sciences. *Reviews of Modern Physics, 91*(4), 045002. https://doi.org/10.1103/RevModPhys.91.045002

[24] Rupp, M., *et al.* (2018). Guest editorial: Special topic on data-enabled theoretical chemistry. *The Journal of Chemical Physics, 148*(24). https://doi.org/10.1063/1.5043213

[25] Chen, J., *et al.* (2025). Hardware-efficient quantum error correction via concatenated bosonic cat qubits. *Nature.* https://doi.org/10.1038/s41586-025-08642-7

[26] Kjaergaard, M., Schwartz, M. E., Braumüller, J., Krantz, P., Wang, J. I.-J., Gustavsson, S., & Oliver, W. D. (2020). Superconducting qubits: Current state of play. *Annual Review of Condensed Matter Physics, 11,* 369–395. https://doi.org/10.1146/annurev-conmatphys-031119-050605

[27] Gambetta, J., *et al.* (2024). IBM Quantum breaks the 100-qubit processor barrier | IBM Quantum Computing Blog — ibm.com. *IBM Quantum Blog.* https://www.ibm.com/quantum/blog/127-qubit-quantum-processor-eagle

[28] *Quantum Computer*. (n.d.). Google Quantum AI. https://quantumai.google/quantumcomputer

[29] *Rigetti QCS*. (n.d.). https://qcs.rigetti.com/qpus

[30] *Accelerating the advent of the quantum computer | QuantWare*. (n.d.). https://www.quantware.com/

[31] Chen, J.-S., *et al.* (2024). Benchmarking a trapped-ion quantum computer with 30 qubits. *Quantum, 8*, 1516. https://doi.org/10.22331/q-2024-11-07-1516

[32] Ye, M., & Delfosse, N. (2025). Quantum error correction for long chains of trapped ions. *arXiv preprint arXiv:2503.22071.* https://arxiv.org/abs/2503.22071

[33] *Quantinuum | Accelerating Quantum Computing*. (n.d.). https://www.quantinuum.com/

[34] *IonQ | Trapped Ion Quantum Computing*. (n.d.). IonQ. https://ionq.com/

[35] Alpine Quantum Technologies. (2024). *Home - AQT - Alpine Quantum Technologies — aqt.eu*. https://www.aqt.eu/
[36] *Oxford Ionics | High Performance Quantum Computing*. (n.d.). Oxford Ionics. https://www.oxionics.com/
[37] Le Régent, F.-M. (2025). *Awesome quantum computing experiments: Benchmarking experimental progress towards fault-tolerant quantum computation.* arXiv. https://arxiv.org/abs/2507.03678
[38] *Billions of qubits on a single chip | Diraq*. (n.d.). Diraq. https://diraq.com/
[39] Quobly, shaping a better digital future. (2025, July 16). *Quobly, shaping a better digital future*. Quobly, Shaping a Better Digital Future. https://quobly.io/
[40] *IonQ | Trapped Ion Quantum Computing*. (n.d.). IonQ. https://ionq.com/
[41] Xanadu. (2024). *Xanadu | Welcome to Xanadu — xanadu.ai*. https://www.xanadu.ai/
[42] ORCA Computing. (2025, July 23). *Home | ORCA Computing*. https://orcacomputing.com/
[43] *Quantum Computing Inc*. (n.d.). https://quantumcomputinginc.com/
[44] *PsiQuantum - Building the world's first useful quantum computer*. (2025, June 17). PsiQuantum. https://www.psiquantum.com/
[45] Pasqal. (2025). *Quantum processing units – product brochure.* https://www.pasqal.com/wp-content/uploads/2025/06/2025-Pasqal-Quantum-Computing-Processor-Brochure.pdf
[46] Saffman, M. (2025). Quantum computing with atomic qubit arrays: Progress and perspectives. *arXiv preprint arXiv:2505.11218.* https://arxiv.org/abs/2505.11218
[47] Pasqal. (2025, June 16). *Home - Pasqal*. https://www.pasqal.com/
[48] Atom Computing. (2025, July 16). *Home - Atom Computing*. https://atom-computing.com/
[49] Infleqtion. (2025, July 23). *Home - infleqtion*. Infleqtion. https://infleqtion.com/
[50] *Quantum Computing with Neutral Atoms | QuEra*. (n.d.). https://www.quera.com/
[51] Smalley, I., & Schneider, J. (2024). What is quantum computing? | IBM — ibm.com. *IBM*. https://www.ibm.com/topics/quantumcomputing#:~:text=Quantum%20computers%20use%20circuits%20with,individual%20units%20of%20quantum%20information
[52] Bal, M., *et al.* (2024). Systematic improvements in transmon qubit coherence enabled by niobium surface encapsulation. *npj Quantum Information, 10*(1). https://doi.org/10.1038/s41534-024-00840-x
[53] Holevo, A. S. (1973). Bounds for the quantity of information transmitted by a quantum communication channel. *Problemy Peredachi Informatsii, 9*(3), 3–11. https://doi.org/10.22331/q-2023-04-13-978
[54] James, G., *et al.* (2023). *An introduction to statistical learning: With applications in Python*. Springer Nature. https://doi.org/10.1007/978-3-031-38747-0
[55] Jadhav, A., *et al.* (2023). Quantum machine learning: Scope for real-world problems. *Procedia Computer Science, 218*, 2612–2625. https://doi.org/10.1016/j.procs.2023.01.235
[56] Bshouty, N. H., & Jackson, J. C. (1995). Learning DNF over the uniform distribution using a quantum example oracle. In *Proceedings of the Eighth Annual Conference on Computational Learning Theory* (pp. 118–127). https://doi.org/10.1137/S0097539795293123
[57] Cross, A. W., *et al.* (2015). Quantum learning robust against noise. *Physical Review A, 92*, 012327. https://doi.org/10.1103/PhysRevA.92.012327
[58] Rath, M., & Date, H. (2024). Quantum data encoding: A comparative analysis of classical-to-quantum mapping techniques and their impact on machine learning accuracy. *EPJ Quantum Technology, 11*(1). https://doi.org/10.1140/epjqt/s40507-024-00285-3
[59] Bhattaraprot B., & Smanchat, S. (2023). Hybrid quantum encoding: Combining amplitude and basis encoding for enhanced data storage and processing in quantum computing. In *2023 20th International Joint Conference on Computer Science and Software Engineering (JCSSE)* (pp. 512–516). IEEE. https://doi.org/10.1109/JCSSE58229.2023.10201947
[60] Rath, M., *et al.* (2023). Quantum-assisted simulation: A framework for designing machine learning models in the quantum computing domain. *arXiv preprint arXiv:2311.10363.* https://arxiv.org/abs/2311.10363
[61] Aleksandrowicz, G., Alexander, T., Barkoutsos, P., Bello, L., Ben-Haim, Y., Bucher, D., Cabrera-Hernández, F. J., Carballo-Franquis, J., Chen, A., Chen, C.-F., *et al.* (2019). *Qiskit: An open-source framework for quantum computing*. Zenodo. https://zenodo.org/records/2562111
[62] *Cirq*. (n.d.). Google Quantum AI. https://quantumai.google/cirq
[63] Svore, K., *et al.* (2018). Q# enabling scalable quantum computing and development with a high-level DSL. In *Proceedings of the Real World Domain Specific Languages Workshop 2018* (pp. 1–10). https://doi.org/10.1145/3183895.3183901
[64] Smith, R. S., *et al.* (2016). A practical quantum instruction set architecture. *arXiv preprint arXiv:1608.03355.* https://doi.org/10.48550/arXiv.1608.03355
[65] Broughton, M., *et al.* (2020). TensorFlow Quantum: A software framework for quantum machine learning. *arXiv preprint arXiv:2003.02989.* https://arxiv.org/abs/2003.02989
[66] Bergholm, V., Izaac, J., Schuld, M., Gogolin, C., Ahmed, S., Ajith, V., Alam, M. S., Alonso-Linaje, G., AkashNarayanan, B., Asadi, A., *et al.* (2018). *Pennylane: Automatic differentiation of hybrid quantum-classical computations*. *arXiv preprint arXiv:1811.04968.* https://doi.org/10.48550/arXiv.1811.04968
[67] D-Wave Systems. (2024). *D-Wave Ocean Software Documentation; Ocean Documentation 8.1.0 documentation*. https://docs.ocean.dwavesys.com/en/stable/
[68] *OpenFermion | Google Quantum AI.* (n.d.). Google Quantum AI. https://quantumai.google/openfermion
[69] *Cloud Quantum Computing Service - Amazon Braket - AWS*. (n.d.). Amazon Web Services, Inc. https://aws.amazon.com/braket/
[70] William A. Fedak and Jeffrey J. Prentis. The 1925 born and jordan paper "on quantum mechanics". American Journal of Physics,

77(2):128–139, February 2009. ISSN 0002-9505. DOI: 10.1119/1.3009634. URL https://doi.org/10.1119/1.3009634.
[71] Trevor Hastie, Robert Tibshirani, Jerome Friedman, Trevor Hastie, Robert Tibshirani, and Jerome Friedman. High-dimensional problems: p n. The Elements of Statistical Learning: Data Mining, Inference, and Pre-diction, pages 649–698, 2009. URL "https:15//www.sas.upenn.edu/~fdiebold/ NoHesitations/BookAdvanced.pdf".
[72] M. Kjaergaard, M.E. Schwartz,J. Braumüller, P. Krantz, J.I.-J. Wang,S. Gustavsson, and W.D. Oliver. Superconducting qubits: Current state of play. Annual Review of Condensed Matter Physics,11:369–395, 2020. DOI: 10.1146/annurevconmatphys-031119-050605.
[73] P. Sales Rodriguez et al. Experimental demonstration of logical magic state distillation. Nature, 2025. DOI: 10.1038/s41586-025-09367-3. URL https://doi.org/10.1038/s41586-025-09367-3.
[74] Peter W Shor. Algorithms for quantum computation: discrete logarithms and factoring. In Proceedings 35th annual symposium on foundations of computer science, pages 124–134. IEEE, 1994. DOI:https://doi.org/10.1109/SFCS.1994.365700.